\documentstyle[12pt,epsfig,color]{article}
\newcommand{\spacing}[1]{\renewcommand{\baselinestretch}{#1}\large\normalsize}
\spacing{1.5}
\begin{document}

\begin{center}
{\large\bf{Anisotropic flow: A case for Bottomonia}} \\
\bigskip

{\small
Debasish Das$^{a}$\footnote{email : debasish.das@saha.ac.in,dev.deba@gmail.com(corresponding author)}, 
Nirupam Dutta$^{b}$\footnote{email :nirupamdu@gmail.com}
\medskip

$^a$Saha Institute of Nuclear Physics, HBNI, 1/AF, Bidhannagar, Kolkata 700064, India\\
$^b$School of Physical Sciences, National Institute of Science Education and Research Bhubaneswar,
P.O. Jatni, Khurda 752050, Odisha, India\\
}

\end{center}
\date{\today}
\begin{abstract}

Studies of $J/$$\psi$ $v_2$ at RHIC and LHC energies have provided 
important elements towards the understanding on the production mechanisms and on the thermalization of charm quarks. 
Bottomonia has an advantage since it is a cleaner probe. A brief discussion has been provided for 
$\Upsilon(1S)$ $v_2$, which can become the new probe for QGP, including the 
necessity of studies for small systems.

\end{abstract}

\section{\bf{Introduction}}

The calculations based on statistical QCD predict that strongly interacting systems at 
very high energy density and/or temperature~\cite{Wilczek:1999ym,Berges:1998nf} 
are composed of weakly interacting quarks and gluons due to asymptotic 
freedom and Debye screening of colour charge~\cite{Satz:1983jp}. A thermalized 
system where the properties of the system are governed by the quarks and gluons 
degrees of freedom is called the ${\it Quark~-Gluon~Plasma}~(QGP)$~\cite{Shuryak:1980tp}.
The coloured quarks under such extreme conditions are no longer confined to 
hadrons but can roam around the entire system. The hadronic system would then 
dissolve into its constituents, quarks and gluons such that the bulk 
properties of that hadronic system would be governed by these degrees of 
freedom. This is the primary reason for the present heavy ion program ongoing 
at the Relativistic Heavy-Ion Collider (RHIC, at Brookhaven National Laboratory in New 
York) and at the Large Hadron Collider (LHC, at the European Organization for Nuclear Research in Geneva). 
Several probes to study the properties of QGP have been formulated and proposed.

Among those probes, the azimuthal anisotropy or the elliptic flow ($v_2$) of the 
produced particles have been shown to be sensitive to the initial condition and the equation of state (EoS) 
of the evolving matter formed in relativistic heavy ion collisions~\cite{Voloshin:2008dg,Petersen:2014yqa,Dusling:2015gta}.
An anisotropic spatial configuration is found in non-central collisions 
which together with the interactions among the constituents develop pressure gradients 
of different magnitude along different spatial directions. With such expansion 
the spatial anisotropy reduces and the momentum space anisotropy 
builds up rapidly. A Fourier expansion of the invariant triple differential distributions 
are used for understanding the patterns of anisotropic flow, which can be defined as:

\begin{equation}
E \frac{{\mathrm d}^3 N}{{\mathrm d}^3 {\bf p}} =
\frac{1}{2\pi} \frac {{\mathrm d}^2N}{p_{\mathrm T}{\mathrm d}p_{\mathrm T}{\mathrm d} y}
  \left(1 + 2\sum_{n}  v_{n}(p_{\mathrm T},y) \cos[n(\varphi-\Psi_{n})]\right),
\label{invariantyield}
\end{equation}
where $E$ is the energy of the particle, $p$ the momentum, $p_{\rm T}$
the transverse momentum, $\varphi$ the azimuthal angle, $y$ the rapidity,
and $\Psi$ the reaction plane angle. The sine terms in such an
expansion vanish due to the reflection symmetry with respect to the
reaction plane. The term $v_2$ is a  measure of this momentum space anisotropy which is defined as:
$v_{2} =\langle\cos(2(\varphi-\Psi))\rangle=
\langle{p_x^2-p_y^2}\rangle/\langle{p_x^2+p_y^2}\rangle$,
where $p_x$ and $p_y$ are the $x$ and $y$ component of the particle momenta. As explained 
in Eq.~\ref{invariantyield}, $\varphi$ is the azimuthal angle of the produced particles 
and $\Psi$ is the angle subtended by the plane containing the beam axis and impact 
parameter with $x$-direction~\cite{Voloshin:2008dg}. Comparison of $v_{2}$ results from experiments 
with those calculated using relativistic hydrodynamic and transport approaches 
have lead to several important insights towards the understanding of 
QGP~\cite{Ollitrault:1992bk,Alver:2010gr,Teaney:2010vd,Borghini:2005kd,Gardim:2011xv,Teaney:2012ke,Bravina:2013ora}. 
The paper is organised to start with a brief introduction of flow in general and in section 2 
we have a brief survey of existing models and experimental results on charmonia flow and various 
complications which makes the issue very diificult for any kind of convincing conclusion. Bottomonia 
on the otherhand is free from such constraints. Section 3 is devoted for the discussion of why bottomonia 
is more suitable for measuring the anisotropic flow and relevant physics aspects that contribute to the bottomonia 
$v_2$ are discussed. Finally, we summarise by pointing out the challenges that lie ahead.


\section{\bf{Charmonia Flow}}
\label{sec:charm_flow_sec}

Heavy quarks and quarkonia have been considered as celebrated probes for various properties of QGP since 
decades for their unique role in diagnosing the hot dense QCD medium~\cite{Matsui:1986dk,Kluberg:2005yh,Brambilla:2004wf}. 
Charmonia and bottomonia, due to large heavy quark masses are very much expected to be formed early 
in the collisions through initial hard scattering processes~\cite{Das:2011bj}. They witness 
the initial anisotropy of the medium produced through high energy nucleus-nucleus collisions which can be realised 
in the final measurement of quarkonium yield in the experiments. Thus the measurement of $v_2$ for heavy quarkonia 
can provide essential information on the properties of the strongly-interacting system formed in the pristine 
stages of relativistic heavy-ion collisions~\cite{Snellings:2011sz}. However, in this context, survival of quarkonia 
is an persistent issue to deal with. More precisely, the production of heavy quarkonia in various stages of 
medium evolution and their destruction in the medium must be addressed appropriately. At the same time, 
this makes the whole issue of anisotropic flow of quarkonia a recondite subject. 

Firstly, initially produced charmonia and bottomonia have to face the high energy density medium which offers 
color screening~\cite{Matsui:1986dk} to various bound states and dissociates them but $J/$$\psi$ and $\Upsilon(1S)$ 
are believed to be surviving still in the medium, as the dissociation temperature for them are much higher than QCD 
critical temperature. Results from Lattice QCD have shown that $J/$$\psi$ survives even above $T_{c}$ and calculations 
show that they would survive, as spatially compact $c\overline{c}$(quasi-bound) states below 2$T_{c}$\cite{Iida:2006mv}. 
Comprehensive experimental results at SPS~\cite{Kluberg:2005yh,Abreu:2000ni} (including feeddown from other less 
bound resonances like $\psi(2S)$  and $\chi_{c}$) and at RHIC for $J/$$\psi$ production in nucleus-nucleus(AA) 
collisions, we observe that the strongly bound $J/$$\psi$ ground state is suppressed~\cite{physreptvogt1,Adler:2004ta,
Abelev:2009qaa,Adare:2006ns}.

Secondly, there is scattering with the thermal partons~\cite{Song:2015bja,Song:2014qoa,Liu:2009nb,Zhao:2011cv} or gluo-dissociation process 
which also can dissociate various quarkonium states. Therefore, the in-medium cross section for elastic scattering is 
important to know. Survived quarkonia escaping in the direction where the in medium path length is less will have a 
better chance of survival than one along where it experiences a longer path length. The primordial $J/$$\psi$, 
emerges in-plane while passing through the medium and those travel out-of-plane travel more and thus resulting 
in a small azimuthal anisotropy for the $J/$$\psi$. But this is not the whole story as there are recombination 
and regeneration issues related to in-medium quarkonia production in heavy ion collisions as they can be 
dominated by the regeneration from thermalised heavy quarks and antiquarks in QGP~\cite{Song:2015bja,Song:2014qoa,Chen:2016mhl}. 
Though the less bound states does not survive color screening, they can be created late in the medium through 
recombination which we have mentioned. As a result we can expect anisotropic flow for these states as well. 
On a very different analogy, less bound quarkonium states can be present in the medium if quarkonia are formed 
late in the medium through the evolution of a superposed states of different quarkonia species and they may have 
helped to create other excited states in the medium as it cools below the dissociation temperature of those 
states. However at LHC energies, the $J/$$\psi$ production, could be even enhanced more due to the coalescence 
of uncorrelated $c\bar{c}$ pairs in the medium~\cite{BraunMunzinger:2000px,Andronic:2011yq}. Quarkonia formation in the 
statistical model~\cite{Song:2010er}through the recombination of un-correlated $c\overline{c}$ or $b\overline{b}$ 
pair may lead to a negative $v_2$ at low $p_T$ as the medium is boosted with a certain velocity. Regenerated $J/$$\psi$ 
receives the elliptic flow of the charm quarks when they thermalize in QGP resulting in 
large $J/$$\psi$ $v_2$~\cite{Zhao:2012gc} but on the other side, 
some exploratory studies~\cite{Borghini:2011yq} show that the thermalization or 
quasi-equilibration time of quarkonia is almost comparable to the life time of QGP. In that case this 
is not beyond doubt  that quarkonia will be affected through the hydrodynamical evolution of the thermalised medium. 

In this context, one more important aspect is the temporal evolution of the fireball which leads 
to non-adiabatic evolution~\cite{Dutta:2012nw} of quarkonium states in the medium. The temperature of the 
medium falls off rapidly which induce a very rapid change in the in-medium quark anti-quark potential. 
This essentially makes the heavy quark potential as a time dependent quantity and due to this 
time dependence, quarkonia evolve non-adiabatically in the medium. As a result, there is 
always reshuffling among all possible states. In the reshuffling picture, the 
survival probability of the ground state or the transition probabilities to different excited 
states depends on how much time quarkonia have spent in the medium~\cite{Dutta:2012nw}. As the escape time 
is different in different direction, the effect of non-adiabaticity should be contained in the anisotropic 
flow of $J/\Psi$ and other excited states. This scenario has not been considered so far in the investigations 
we have discussed above but the effect of non-adiabatic evolution on flow of $J/$$\psi$ can be markedly 
different from earlier predictions.

In the experimental side, measurement for $v_2$ of $J/$$\psi$ has been extensively performed in past years. 
At RHIC, the results from STAR Collaboration on $J/$$\psi$ $v_2$ in Au+Au collisions at $\sqrt{s_{\rm NN}} = 200$ GeV 
are consistent with zero, within the large measurement uncertainties~\cite{Adamczyk:2012pw}. Also (preliminary) 
measurements from the PHENIX Collaboration in Au+Au collisions at $\sqrt{s_{\rm NN}} = 200$ GeV~\cite{Silvestre:2008tw} 
show similar results.Positive $J/$$\psi$ $v_2$ was observed by ALICE Collaboration in semi-central 
Pb+Pb collisions at $\sqrt{s_{\rm NN}} = 2.76$ TeV with a $2.7\sigma$ significance for the inclusive $J/$$\psi$ with 
$2 < p_{\rm T} < 6$~GeV/$c$ at forward rapidity~\cite{ALICE:2013xna}.The CMS Collaboration also reported a positive 
$v_2$ for prompt $J/$$\psi$ at high $p_{\rm T}$ at mid-rapidity~\cite{Khachatryan:2016ypw}. Better measurements of 
the $J/$$\psi$ $v_2$ in Pb+Pb collisions at $\sqrt{s_{\rm NN}} = 5.02$ TeV~\cite{Acharya:2017tgv} with the ALICE detector are studied 
at mid-rapidity ($|y| < 0.9$) in the di-electron decay channel and compared with forward rapidity ($2.5<y<4.0$) 
in the dimuon channel, both down to zero transverse momentum. A positive $v_2$ is again observed in the transverse 
momentum range $2 < p_{\rm T} < 8$ GeV/$c$ in the three centrality classes. This affirms well with the results at 
$\sqrt{s_{\rm NN}} = 2.76$ TeV~\cite{ALICE:2013xna} in semi-central collisions. At mid-rapidity, the $J/$$\psi$ $v_2$ 
is explored as a function of transverse momentum in semi-central collisions and found comparable with the 
measurements at forward rapidity for Pb+Pb collisions at $\sqrt{s_{\rm NN}}$=5.02 TeV~\cite{Acharya:2017tgv}.

At this juncture, can we really say something convincing about the initial state anisotropy of the medium through the 
study of $v_2$ of quarkonia? We have realised from our discussion so far that picture for charmonia is quite messy 
as because many effects in various phases of the hot medium seems to be contributing appreciably in the prediction 
of $v_2$. Various models have been employed to explain the experimental results discussed above but have faced difficulties 
to accomodate all the different mechanism involved in the process. On the other hand, 
the scenario for bottomonia is comparatively simple. For bottomonia, the possibility of recombination through 
uncorrelated $b$, $\bar{b}$ is less expected as there are not much of them in the medium in compare to $c$, $\bar{c}$ 
and the usual expectation is that the leading effect is suppression and then other contributions come from elastic 
scatterings as well as the possible excitation de-excitation of bottom-antibottom bound states. These two effects 
might be sub leading for bottomonia but never the less one must incorporate the reshuffling of all possible 
bottomonium states into consideration. In the next section we will discuss this concept in more detail.  


\section{Bottomonia Flow}
\label{sec:bottom_flow_sec}

Due to regeneration, the picture of $J/$$\psi$  suppression or its absence
may not lead us towards the final answer for QGP signature at collider energies.
At this point the study of bottomonia ($\Upsilon(1S)$, $\Upsilon(2S)$, $\Upsilon(3S)$) candidates assumes
importance. Since the $\Upsilon$ is smaller than $J/$$\psi$, its absorption cross section should
be smaller. In terms of AA collisions, the $\Upsilon$ is expected to be dissociated at a
higher temperature than all the other quarkonium states, thus proving to be a more
effective thermometer of the system~\cite{Kisslinger:2012tu}. 
The high collision energies and luminosities have made the possibility of
studying bottomonium production in heavy-ion collisions recently at the RHIC and 
LHC energy regimes~\cite{Adamczyk:2013poh,Adare:2014hje,Chatrchyan:2011pe,
Chatrchyan:2012lxa,Abelev:2014nua,Sirunyan:2017lzi}.

When compared with the $J/$$\psi$ case, the probability for the $\Upsilon$ states to be regenerated in the medium 
is small due to the lower production cross section of the $b\overline{b}$ pairs~\cite{Grandchamp:2005yw,Emerick:2011xu}. 
But the feed-down from higher mass bottomonia (between $40\%$ and $50\%$ for $\Upsilon$(1S)~\cite{Abelev:2014nua}) 
complicates data interpretation. The regeneration is found to be negligible for the $\Upsilon(1S)$ but maybe 
possibly more important for the $\Upsilon(2S)$. Hence, the flow pattern of bottomonia will generally not acquire 
a contribution due to the flow of $b$ and $\bar{b}$ ``parents'', which should result in a bottomonium $v_2$ smaller 
than that of $J/\psi$. This expectation is supported by the calculations of elliptic flow reported in Ref.~\cite{Du:2017qkv}. 
Due to the large mass of bottomonia, elastic collisions on the medium constituents are expected to be inefficient 
at changing the direction of bottomonia propagation in the medium. Even at energy densities equivalent to 
temperatures of order $T\sim 4T_c$, the medium constituents have transverse momenta $p_{\rm T}$ (close to 3T) 
$\sim$ 2 GeV which is  a factor of 5 smaller than $M_\Upsilon$. Thus for making more breakthrough a clear 
notion of elastic cross-section for bottomonia colliding on QGP is required for model estimates. Besides 
that, the suppression due to QGP must be disentangled from Cold Nuclear Matter (CNM) effects (such as 
nuclear modification of the parton distribution functions or break-up of the quarkonium state in CNM) 
which, as of now, are not accurately known at RHIC energies~\cite{Adamczyk:2013poh}.

But still bottomonia remains attractive, since their evolution is apparently more straightforward to 
describe than that of charmonia. After creation the bottomonia will either propagate through the hot 
environment in an almost free way, or get destroyed. In the latter case, the probability for their 
disappearance is controlled by their in-medium path length and by the energy density of the medium 
they encounter. While expecting $\Upsilon$s ``flow'' or rather in the sense that their emission pattern 
is correlated to the initial geometry of the collision, we would also need to assign a ``survival'' or 
in other words an ``escape mechanism''.  Accordingly, the anisotropic flow of bottomonia can be explained 
by an escape mechanism from a spatially asymmetric medium~\cite{Borghini:2010hy,He:2015hfa,Romatschke:2015dha} 
as was originally proposed for $J/\psi$~\cite{Wang:2002ck}. For instance, bottomonia should have a 
higher survival probability when they propagate along the direction of the impact parameter than 
perpendicular to to the reaction plane, leading to a positive $v_2$. 

The spatial asymmetries probed by the survived bottomonia, will be early-time asymmetries. This opens the 
possibility that the higher-harmonic eccentricities effectively seen by the bottomonia should still be 
close to those in the initial state, instead of having been damped by the dissipative effects in 
the medium evolution~\cite{Staig:2011wj}. Measurements of the integrated anisotropic 
flow of bottomonia could thus give an alternative access to the event-by-event fluctuations of the initial 
geometry, somewhat different from the information given by the flow of charged particles, especially 
regarding the higher anisotropies. However such argument also holds for the fraction of ``primitive'' 
$J/\psi$ that will survive suppression. But its effect is masked by the contribution to flow of $J/\psi$ 
from regeneration. In that respect, bottomonia appear to be a cleaner probe than charmonia as far as 
initial fluctuations are concerned.

The above discussion is based on the traditional idea according to which a $b\bar{b}$ pair initially in the 
$\Upsilon(1S)$ state, or any other, remains in that same state during the medium evolution. Here we feel 
the necessity of discussing reshuffling picture which so far in the context of quarkonia flow has not been 
considered. It was put forward in the context of real time evolution of heavy quarkonia states in the open 
quantum system framework~\cite{Borghini:2011yq,Dutta:2012nw,Borghini:2011ms,Akamatsu:2011se,Brambilla:2016wgg,Blaizot:2017ypk}. 
Reshuffling picture becomes relevant here because the quark-antiquark potential in the medium is time 
dependent for two reasons. Firstly, the medium produced in heavy ion collisions cools off very rapidly 
and as a result the heavy quark potential changes with time. In a time varying potential, bottomonia 
evolves non-diabatically~\cite{Dutta:2012nw} and consequently, the initially survived ground state evolves 
as a superposition of all possible states. Therefore, the ground state starts projecting itself to 
other less bound excited states. Furthermore, if we consider the movement of bottomonia inside the 
medium, the potential changes as they move from one region to another region of the anisotropic medium. 
The rate of change of quarkonium potential $V$ due to temporal evolution of the medium and the motion 
of quarkonia can be expressed as,
\begin{equation}
\frac{dV}{dt} = (\vec v.\vec \nabla )V+\frac{\partial V}{\partial t}
\end{equation}
The probability of transitions among different states are proportional 
to $\frac{\langle \psi_m|\frac{dV}{dt}|\psi_n\rangle}{|{E_m-E_n}|^{2}}$, where $|\psi_n\rangle$ 
and $E_n$ signifies the $n$th eigenstate  and energy of bottomonia respectively. Quarkonia are described 
outside the medium through vacuum Cornell potential which is different from the one at the centre of the medium. 
Hence, bottomonia escaping along the direction where the in-medium path length is smaller have a bigger 
change in the potential compared to one along which the path length is bigger. Hence the transition 
probabilities of other excited states for those traveling less are expected to be higher than for those who are travelling 
along the bigger path length. As a result, we can expect a positive $v_2$ for other excited states. 
We also notice that the transition probability to different excited states from $\Upsilon(1S)$ is 
inversely proportional to the energy gap between the two states. Hence, for those excited states 
which are energetically closer to the groud state are expected to be created more than the distant states. 
This fact may result in a observation of prominent positive $v_2$ for them and 
we might not witness $v_2$ for distant excited states as they cannot be created much 
through the non-adiabtic evolution.

Transitions from the $\Upsilon(1S)$ to a less bound state will result in a smaller survival 
probability of it, and thus can be accounted for an effective increase of the absorption cross section. 
However, the suppression can then take place at later times, i.e. at lower temperatures. Conversely, 
transitions from excited to the ground state, if they happen quickly enough, may protect the bound 
state from destruction. In either case, model studies are needed to quantify the resulting influence 
on the anisotropic flow of bottomonia. Nevertheless, if the flow coefficients of various bottomonium 
states happen to be similar, then it would be a strong hint that reshuffling does occur during the 
medium evolution. Without reshuffling, the surviving excited states would mostly originate from the 
edges of the medium, in contrast to the $\Upsilon(1S)$, so that they would probe different geometries 
and thus acquire different anisotropic flow patterns.

As discussed before the effects related to the presence of CNM can also modify the production of 
quarkonia in AA collisions. Cold nuclear matter effects can be studied in proton-nucleus (pA) collisions, 
where the QGP is not expected to be formed. Due to the larger mass of the bottomonium states compared to the 
charmonium ones, the measurement of bottomonia production in proton-nucleus 
collisions~\cite{Aaij:2014mza,Abelev:2014oea,Chatrchyan:2013nza} allows a study of cold nuclear 
matter effects in a different kinematic regime, therefore complementing the J/$\psi$ 
studies~\cite{Aaij:2013zxa,Abelev:2013yxa,Aad:2015ddl}. For ``smaller systems'' like pA we 
have less deeply bound bottomonia states and thus a comparatively larger chance to escape.
This means that more states become measurable, which is a positive feature. 
On the other hand, it also means that the escape mechanism which underlies 
the anisotropic flow of bottomonia may become largely ineffective, in particular for the $\Upsilon(1S)$. 
Accordingly, the measurement of a sizable flow for $\Upsilon(1S)$ in small systems would 
probably hint at the importance of initial-state correlations as advocated for instance in Ref.~\cite{Dusling:2017dqg}. 
In particular, a parallel behavior of the flow coefficients of bottomonia and charged particles 
across centralities would be seen as a strong case against the ``hydrodynamical'' explanation of 
charged-particle flow in small systems. Note, however, that the mechanism underlying heavy quark 
production may differ from that responsible for the bulk of particles, so that our qualitative 
statement needs support from explicit calculations of the multiparticle correlations involving bottomonia. 
Recent presence of flow-like signals in collisions of small systems as in p-Pb and in high-multiplicity 
p-p collisions at the LHC are shown in Ref.~\cite{Nagle:2018nvi} which presents an overview 
of the experimental results.

\section{Summary and Outlook}

The $J/$$\psi$ elliptic flow has helped us to understand the charm quark thermalization 
in the medium. However, it is also clear that the picture provided by $J/$$\psi$ needs to be 
complemented with bottomonia($\Upsilon$s) since there is a considerable aspect of 
regeneration which is small for bottomonia candidates. Regarding experimental studies at 
RHIC the measurements may still become challenging due to the constraints on 
$\Upsilon(1S)$ counts in top Au+Au energies~\cite{Adamczyk:2013poh,Adare:2014hje} keeping 
the options left for coming large Pb+Pb data-sets at top LHC energies in 2018.

\bibliography{apssamp}

\end{document}